\documentclass[aps,floats,amssymb,prl,twocolumn,showpacs,superscriptaddress]{revtex4}
\usepackage{calc}
\usepackage{psfrag}
\usepackage{graphicx}
\begin{document}
\def \beq{\begin{equation}}
\def \eeq{\end{equation}}

\begin{abstract}
We study by exact diagonalization, in the lowest Landau level approximation,
the Coulomb interaction problem of $N = 4$ and $N = 6$ quantum dot in the
limit of zero Zeeman coupling. We find that meron excitations constitute
the lowest lying states of the quantum dots. This is based on a mapping
between the excitations of the dot and states of
the Haldane-Shastry spin chain.
\end{abstract}


\title{Fractionalization into merons in quantum dots}
\author{A. Petkovi\'{c}}

\affiliation{Institute of Physics, P.O.Box 68, 11080 Belgrade,
Serbia}
\affiliation{Institut f\"ur Theoretische Physik,
Universit\"at zu K\"oln, 50937 K\"oln, Germany}

\author{M.V. Milovanovi\'{c}}

\affiliation{Institute of Physics, P.O.Box 68, 11080 Belgrade,
Serbia}

\date{\today}
\maketitle \vskip2pc
\narrowtext

{\em Introduction} Quasiparticles with fractional charge and
fractional statistics make a hallmark of the fractional quantum Hall (QH)
effect \cite{laug,halp}. They are usually found in spin polarized systems, but they
can be found also in the systems where Zeeman energy is negligible
and spin degree of freedom plays a role. In these systems quasiparticles,
which carry charge, can be identified with topological objects,
special configurations of spin in space. An example of this is
skyrmion, the topological object well known from classical
ferromagnetism \cite{rraj}, which is also a quasiparticle in the $\nu = 1$
 QH system \cite{so}. We will show in this letter that as the size of the $\nu = 1$
system shrinks to quantum dot (QD) another quasiparticles of a
topological nature, merons \cite{oref}, enter the stage and constitute
the lowest lying excitations.

Each meron is a half of skyrmion \cite{moo}. If in the center of skyrmion excitation
in 2D plane spin points up it slowly tumbles down to the down configuration
on its circular boundary. On the other hand in the case of meron spin
does not reach the down position on its boundary but it is half way
between the up and down position i.e. it is in the plane and winds
for $2 \pi$ along its boundary; see Fig. 1.

\begin{figure}
\centering
\includegraphics[width=0.8\linewidth]{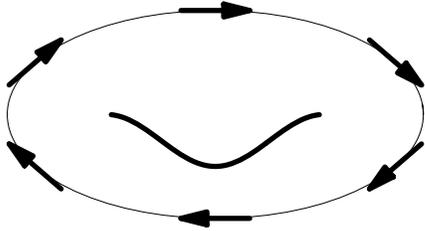}
\caption{Meron quasihole.}\label{meron}
\end{figure}

It is hard to prove existence of a quasiparticle in a finite, small
system. In order to prove the existence of merons in QDs we used
a mapping between seemingly uncorrelated physical systems, 2D QD with
zero Zeeman coupling and 1D Haldane-Shastry (HS) spin chain \cite{hs}. It is not utterly
suprising that it exists and maps the physics of spinons, excitations,
of the HS chain to the one of merons in QD. It is well known that the
edge of QH systems can be mapped to one-dimensional models (charge
excitations to Calogero-Sutherland model \cite{yy}) and this may exist even in small systems like QDs where
the distinction between the bulk and edge is blurred.

{\em Model} We model a quantum dot in the regime of high magnetic
fields in the lowest Landau level (LLL) approximation . Accordingly
the Hamiltonian takes the following form,
\begin{equation}
H = H_{sp} + H_{int},
\end{equation}
where $H_{sp}$ denotes the single particle part, without the Zeeman term
in our case,
$H_{sp} = \hbar [\omega N + (\omega - \frac{1}{2} \omega_{c}) L]$,
where $\omega_{c} = \frac{e B}{m^{*}}$, the cyclotron frequency,
$\omega = \sqrt{\omega_{o}^{2} + \frac{\omega_{c}^{2}}{4}}$ where
$\omega_{o}$ is the frequency of the harmonic confining potential,
$L$ and $N$, the total orbital angular momentum and number of
particles of the dot respectively, and
$H_{int}$ denotes the interaction part,
\begin{equation}
H_{int} = \frac{1}{2}
\sum
a^{\dagger}_{m_{1} \sigma} a^{\dagger}_{m_{2} \sigma'} a_{m_{3}
\sigma'} a_{m_{4} \sigma} \scriptstyle \langle m_{1}|\langle m_{2}|
\textstyle V_{12} \scriptstyle |m_{4}\rangle|m_{3}\rangle,
\end{equation}
where $ a^{\dagger}_{m \sigma} $ and $a_{m \sigma}$ create and
annihilate electron with the spin projection $\sigma$ in the single
particle state of the LLL with angular momentum  $ m \geq 0$,
\begin{equation}
<\mathbf{r}|m> = \frac{1}{\sqrt{2 \pi 2^{m} m!}} r^{m} \exp\{-im
\phi\} \exp\{- \frac{r^{2}}{4}\},
\end{equation}
where $\hbar =1 $ and $2 m^{*} \omega = 1$.
$V_{12}$ is the Coulomb interaction operator, $V_{12} = \frac{e^{2}}{4 \pi \epsilon} \frac{1}{|\vec{r}_{1} - \vec{r}_{2}|}$,
between two particles. As $H_{sp}$ is trivially diagonalized and accounted for, we will refer in the
following to the energies of $H_{int}$ as those of $H$.

{\em HS spin chain}
The HS Hamiltonian is
\begin{equation}
H_{HS} = J (\frac{2\pi}{N})^{2} \sum_{\alpha < \beta}^{N}
\frac{\vec{S}_{\alpha} \vec{S}_{\beta}}{|z_{\alpha} -
z_{\beta}|^{2}},
\end{equation}
where sites $\alpha, \beta = 1, \ldots, N$ are positioned on a unit circle so that
each site coordinate, $ z_{\alpha}$, fulfills $z_{\alpha}^{N} - 1 = 0$, and
$\vec{S}_{\alpha}$'s are spin 1/2 operators.
Any state of the chain can be represented as a function of complex
numbers satisfying $z_{\alpha}^{N} = 1$, and representing sites for
which $S_{\alpha z} = +1/2$. In this way the ground state wave
function, a spin singlet, is a function of $N/2$ complex numbers,
$\Psi(z_{1}, \ldots, z_{N/2}) = \prod_{j < k}^{N/2} (z_{j} -
z_{k})^{2} \prod_{j = 1}^{N/2} z_{j}.$ Spinons, elementary
excitations, are quasiparticles of spin 1/2.
The hallmark of the HS spin chain is the existence of
``supermultiplets", degenerate energy eigenstates of the same spinon
number but different spin. The structure of eigenstates is built on so-called
``fully (spin-)polarized spinon gas" (FPSG) states \cite{hs}
with definite spinon number and maximum spin equal to the half
of the spinon number. Their corresponding, degenerate states can be
found by acting with an operator of the Yangian algebra,
inherent to the model for which the FPSG states are highest weight states.
In the case of two spinons we have two
degenerate states a triplet and a singlet and the latter wavefunction is
$\Psi^{'}(z_{1},\ldots, z_{N/2}) = \prod_{j < k}^{N/2} (z_{j} -
z_{k})^{2} [ 1 - \prod_{j = 1}^{N/2} z_{j}^{2}]$.

{\em Motivation behind Numerical Calculations} We will present our numerical
work later on, but here we will give just a synopsis of what can be established
on the basis of the numerical work and how this can be used to prove the
existence of merons.

In the case of polarized electrons, QD at $\nu =1$ is in a stable
state, so-called maximum density droplet  (MDD) state, where each
angular momentum orbital till its maximum value $N - 1$ is filled.
Therefore its total angular momentum is $M = N (N - 1)/2$. In this
system the existence of vortex (dressed hole) quasiparticle
excitation is firmly established \cite{oak,saar,man}. As we slowly
increase the magnetic field (from the MDD value) a particle-hole
vortex excitation is formed, where the vortex occupies an inner
orbital of the MDD state and at the same time pushes charge outside
and creates a particle on the edge. This process is followed by
gradual increase of $M$. If QD is small enough this description may
persist to the point when the vortex quasiparticle is created at the
center which can be associated with the total increase of $\Delta M
= N$ of the angular momentum as implied by the shift register
counting of Laughlin \cite{qhla}, or, therefore, an increase of one
flux quantum in the magnetic flux through the system. Therefore the
lowest lying excitations of the dot as the magnetic field is
increased can be described through vortex excitations which makes
the vortex an eigenstate of the polarized system.

We will show that a similar process happens with unpolarized electrons,
starting from the MDD configuration \cite{ref}, where the quasiparticle that slowly
sweeps the inner orbitals, pushes charge outside, and finally is created
at the center is meron. Therefore even in this case we will recover usual
quasiparticle-hole phenomenology of a  QH system and therefore
prove the existence of merons. The meron is half of skyrmion, and to skyrmion
an increase of one flux quantum is associated because skyrmions are analogs
of hole (vortex) excitations in the  limit of weak Zeeman coupling \cite{so}. Therefore if we create
a meron at the center we expect half flux quantum increase in the flux through the system
and associated increase of $\Delta M = N/2$ in the total angular momentum.
That should be also period for which we expect appearance of merons of
higher winding number in very small dots; see $N = 4$ example below.

Our findings support that the MDD state, with respect
to its spin content, can be viewed as a condensate of $N$ merons spin
$\frac{1}{2}$ where each two merons pair to one hole (a vortex)
\cite{lc}. We will call these quasiparticle (not quasihole) kind of merons condensate  merons.
We find, while establishing a mapping between the dot and the HS chain
that the ferromagnetically ordered MDD state
corresponds to the $ 2 S = N$, maximum number of spinons, HS $N$ site spin
chain state.
This state, except for the spin degeneracy,
 is a unique state
of the chain. When a quasiparticle, more precisely a  meron quasihole,
enters the dot, the number of condensate merons is reduced, by
creation of a quasiparticle-hole pair, to $N - 2$. We find that as an effect the
succesive quasihole orbital states as it enters a dot can be
associated to  half of (see \cite{ch}) the $N - 2$ spinon sector of the HS $N$ site spin
chain. Therefore merons map to spinons and we are establishing merons as
eigenstates of dot problem \cite{comment}.

\begin{figure}
\centering
\includegraphics[width=0.8\linewidth]{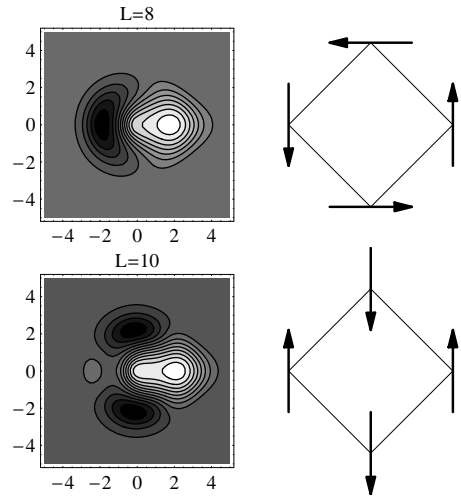}
\caption{Spin-spin correlations along fixed radii of lowest lying, spin-singlet states
 of the $N = 4$ dot. The reference points are on the positive
side of $x$ axis. On the rhs are two possible classical meron configurations
where only the characteristic spin windings along the meron edge are shown.}\label{figN=4}
\end{figure}

{\em N =4 dot} In Fig. 2 we show spin-spin correlations along fixed
radii, of lowest lying, at fixed $L$, spin singlet states. On the
rhs of Fig. 2 shown are two possible types of (classical)
configurations, windings of merons at their edges projected to four
sites. The occurence of black regions correspond to our expectations
where to find antiferromagnetic correlations according to the
classical configurations at some radii, which turn out to be near or
at the maximum (radial) density radius. Similar results already
appeared in \cite{im} but not with emphasis on the $\frac{N}{2}$
periodicity (the $L$ of MDD is at $L = \frac{N(N-1)}{2} = 6$), and
meron interpretation we are pointing out here. And indeed after a
detailed investigation we found also $RVB^{+}$ and $RVB^{-}$ four
site spin states \cite{im} to describe these states, $L = 10$  and
$L = 8$ respectively, but in addition
 we identified these states with the
gound and two spinon lowest lying singlet state of the HS four site spin
chain respectively.
Moreover at $L =8$ we identified, with respect to the lowest lying, a nearly
degenerate $S = 1$ state with the triplet, $z^{2}$ state of the HS, and this
is in the accordance with the mapping to the two degenerate states,
a singlet ($RVB^{-}$) and a triplet ($z^{2}$) of the $N -2 = 2$ spinon sector of the
HS chain. Also at $L = 7$ we found the triplet $z^{3}$ two spinon state of
the chain as expected \cite{hs}.
\begin{figure}
\centering
\includegraphics[width=\linewidth]{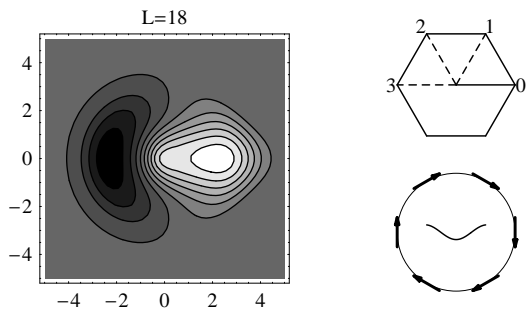}
\caption{The map of the spin-spin correlations of the lowest lying,
spin-singlet state at $L = 18$ of the $N = 6$ dot. In the lower
right corner is the classical meron edge configuration for
comparison. }\label{sl2}
\end{figure}

{\em N =6 dot} Also in the case of the $N = 6$ dot we will recover
meron quasiparticle-hole states as the lowest lying ones for angular
momenta higher than, but still near, the MDD state. First we can
notice in Fig. 3 at $\Delta M = N/2 = 3,\; M = 18$ a meron quasihole
created at the center of the dot. Before reaching that state as the
magnetic field is gradually increased from the MDD state the system
passes through the states at $M = 16$ and $M = 17$ analyzed in Fig.
4. As announced we expect that the $N - 2 = 4$ spinon sector of the
HS chain is associated with the states at $M = 16, 17,$ and $18$.
And indeed just above $L = \frac{N(N - 1)}{2} = 15$ at $L = 16$ and
$L = 17$ we easily identified these HS states as in Fig. 4 where
corresponding states are noted. Plotted are ratios of calculated
real parts of $S^{+} S^{-}$ correlations as functions of radius of
the dot states subtracted by the same ratios for the HS $N = 6$
chain for expected HS states. We take ratios because of different
electron densities at different dot radii. In general we define the
quantities displayed in Fig. 4 as
\begin{equation}
f_{ab} = \frac{Re\langle S^{+}(r, 0) S^{-}(r, a)\rangle}{Re\langle
S^{+}(r, 0) S^{-}(r, b)\rangle} -\frac{Re\langle S^{+}(0)
S^{-}(a)\rangle|_{HS}}{Re\langle S^{+}(0) S^{-}(b)\rangle|_{HS}},
\end{equation}
where $a$ and $b$ take values 1,2, and 3 denoting the angles, with
respect to the positive $x$-axis, represented on the hexagon in the
upper right corner of Fig.3. Other HS states with respect to the
shown do not show the nice confluences of all three lines, $f_{12},
f_{13}$, and $f_{23}$, at their simultaneous value zero at a single
radius; they are completely off and uncorrelated. Similarly we can
define quantities with the imaginary parts of $S^{+} S^{-}$
correlations. Some of them are identically zero, due to the same
real correlation property of HS and QD states, and the rest are
 compatible to and very suggestive of the trend
around the special radius that exists in real parts (Fig. 4).
That again is the behavior we see only for these special HS states.
We find that the HS states' conditions, including  $S_{z} S_{z}$ correlations,
are satisfied at the radius (Fig. 4)slightly beyond the maximum density radius.
\begin{figure}
\centering
\includegraphics[width=\linewidth]{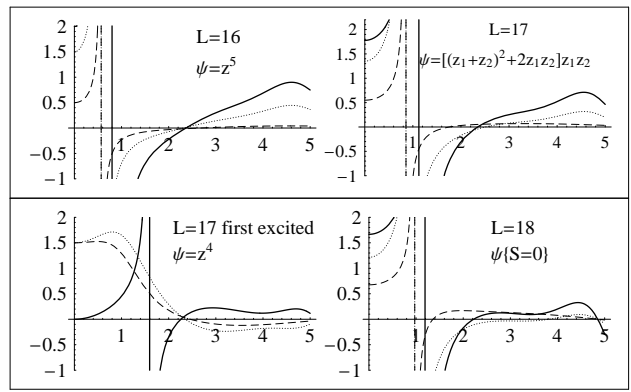}
\caption{Quantities $f_{12}$ - full, $f_{13}$ - dotted, and $f_{23}$ -
dashed, for the definitions see the text, as functions of radius. The
quantities should go to zero, simultaneously if the corresponding HS
state, in the Figure denoted for each angular momentum, for $L = 18$ see
the text, describes
the dot state correlations along a fixed radius. (Asymptotes are
also shown.)}\label{figenergija}
\end{figure}
\begin{figure}
\centering
\includegraphics[width=\linewidth]{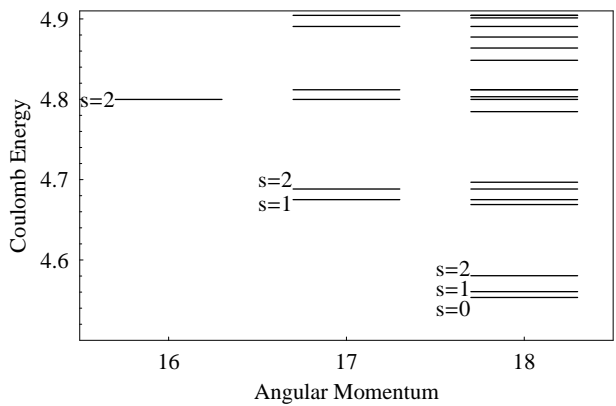}
\caption{The spectrum of the $N =6$ dot as a function of angular
momentum. Lowest energy, closely lying levels correspond to HS
supermultiplets \cite{hs}.}\label{sl1}
\end{figure}

For $L =17$ we found the $S =2$, $z^{4}$ closely lying, first excited
state that belongs to the expected \cite{hs} HS two state degenerate supermultiplet (Figs. 4 and 5.).
We emphasize that although the results of the $N = 4$ case
may be justified or expected because of early (for small L)
Wigner crystal formation \cite{maks}, we checked that for $L = 15 - 18$,
in the $N = 6$ case there is no underlying Wigner crystal configuration \cite{gu}.
The mapping does not require underlying crystal structure, and may
persist in dots with $N$ higher than $6$.

In Fig. 5  we see clearly, beside the $L= 17$ multiplet,
the existence of an additional multiplet at $L = 18$, with three states,
that we expect \cite{hs} if we identify the $S =0$ state as the lowest lying
$S = 0$ state in the 4-spinon sector,
\begin{equation}
 \Psi_{S = 0} = \prod_{i<j}^{M} (z_{i} - z_{j})^{2} \prod_{i}^{M} z_{i} +  15 \{ \prod_{i}^{M} z_{i} +
 \prod_{i}^{M} z_{i}^{3} + \prod_{i}^{M} z_{i}^{5} \},
\end{equation}
where $M = 3$. And indeed we were able to do that, the analysis is
given in Fig. 3, where we see a clear correspondence
between the classical meron edge configuration ($2\pi$ winding of
the spin vector in the plane)
 and the spin-spin correlation map, and in Fig.  4, but
less easily than the lower momentum states. The reason is that we
are already approaching a transition region to the Wigner regime
\cite{bed,har} where
 the Wigner structure
with a pentagon and an electron  in the middle competes
with the hexagonal structure \cite{apmm}. Therefore the transition region
may well consists of liquid states of merons similarly
to what happens in the polarized case with vortices \cite{man}.

{\em Final Remarks} The physics of exchange \cite{so,moo} that
favors smooth variations of spin in space, favors meron solutions
instead of skyrmion ones in small systems, in the limit of zero
Zeeman coupling. Merons distort spin at slower rate than skyrmions,
although acquire also energy due to winding over the space of the
system. This second contribution is suppressed in small systems
\cite{ed} where merons, as we demonstrated here, constitute lowest
lying eigenstates.

 The meron physics
should be  detectable in lateral quantum dots, in which interaction
effects are strong, in their not yet explored regime beyond the MDD
state \cite{ni}. The same conclusions are valid for rapidly rotating
fermi atoms, where there is no Zeeman effect to disguise the
fractionalization into merons, and, therefore, the RVB and spinon
physics implied by the mapping to the HS chain may reveal more
easily.

M.V.M. thanks E. Dobard\v{z}i\'{c} and Z. Radovi\'{c} for previous
collaboration. The work was supported by
Grant No. 141035 of the Serbian Ministry of Science.


\begin{references}
\bibitem{laug} R.B. Laughlin, Phys. Rev. Lett. {\bf 50}, 1395 (1983).
\bibitem{halp} B.I. Halperin, Phys. Rev. Lett. {\bf 52}, 1583 (1984).
\bibitem{rraj} R. Rajaraman, {\it Instantons and solitons}, (North-Holland, Amsterdam, 1987).
\bibitem{so} S.L. Sondhi et al., Phys. Rev. B {\bf 47},
16419 (1993).
\bibitem{oref} D.J. Gross, Nucl. Phys. B {\bf 132}, 439 (1978); I. Affleck, Phys. Rev. Lett.
{\bf 56}, 408 (1986).
\bibitem{moo} K. Moon  et al., Phys. Rev. B {\bf 51}, 5138 (1995).
\bibitem{hs} F.D.M. Haldane, Phys. Rev. Lett. {\bf 60}, 635 (1988); {\bf 66}, 1529 (1991);
B.S. Shastry, Phys. Rev. Lett. {\bf 60}, 639 (1988);
\bibitem{yy} H. Azuma and S. Iso, Phys. Lett. B {\bf 331}, 107
(1994).
\bibitem{oak} J.H. Oaknin et al., Phys. Rev. Lett. {\bf 74}, 5120 (1995).
\bibitem{saar} H. Saarikoski et al., Phys. Rev. Lett. {\bf 93}, 116802 (2004).
\bibitem{man} M. Manninen  et al., Phys. Rev. Lett. {\bf 94}, 106405 (2005).
\bibitem{qhla} R.B. Laughlin in {\it The Quantum Hall Effect}, 2nd ed.,
edited by R.E. Prange and S.M. Girvin (Springer-Verlag, New York, 1990).
\bibitem{ref} Here we use the notion of the MDD state not as of
a maximum possible density state, but as the one in which in the orbital space,
each orbital till the value $N - 1$ is filled by exactly one electron.
\bibitem{lc} We noticed the presence of vortex excitations in the low lying spectrum.
\bibitem{ch} As may be expected only one chirality (angular momentum direction) branch
maps to the states of QD in a magnetic field.
\bibitem{comment} It is not surprising that the mapping we find is to the HS chain. The chain
is a paradign of quantum antiferromagnet fractionalization and its excitations,
spinons, have semionic statistics. Merons as halfs of skyrmions, which at
$\nu = 1$ carry fermionic statistics, are expected to have semionic statistics.
In principle it is hard to talk about statistics in small systems but if
we have signatures of statistics in small chains \cite{hs}, to which mapping is possible,
they should carry information of the expected quasiparticle statistics in QDs.
\bibitem{im} H. Imamura, H Aoki, and P.A. Maksym, Phys. Rev. B {\bf 57}, R4257 (1998); H. Imamura,
P.A. Maksym, and H. Aoki, Physica B 249-251, 214 (1998).
\bibitem{maks} P.A. Maksym, Phys. Rev. B {\bf 53}, 10871 (1996).
\bibitem{gu} The same conclusion was reached in A.D. Guclu and C.J. Umrigar, Phys. Rev. B {\bf 72}, 045309 (2005).
\bibitem{bed} V.M. Bedanov and F.M. Peeters, Phys. Rev. B {\bf 49}, 2667 (1994).
\bibitem{har} A. Harju, S. Siljamaki, and R.M. Nieminen, Phys. Rev. B {\bf 65}, 075309 (2002).

\bibitem{apmm} A. Petkovi\'{c} and M.V. Milovanovi\'{c}, unpublished
\bibitem{ed} E. Dobard\v{z}i\'{c}, M.V. Milovanovi\'{c}, and Z. Radovi\'{c}, unpublished
\bibitem{ni} A  recent study of $N = 5$ electrons in a vertical quantum dot showed partially
polarized states beyond the MDD, Y. Nishi  et al., Phys. Rev. B {\bf
74}, 033306 (2006).
\end{references}
\end{document}